\documentclass[referee]{raa}            

\usepackage{graphicx,times}             

\begin{document}

   \title{VARIABLE STARS OBSERVED WITH THE AST3-1 TELESCOPE FROM DOME A OF ANTARCTICA
}

   \volnopage{Vol.0 (200x) No.0, 000--000}      
   \setcounter{page}{1}          

   \author{Li Gang
      \inst{1}
   \and Fu Jianning
      \inst{1} \footnotemark[1]
   \and Liu Xuanming
      \inst{1}
   }

   \institute{Department of Astronomy, Beijing Normal University,
             Beijing 100875, China;\\
   }

   \footnotetext[1]{Send offprint request to: jnfu@bnu.edu.cn}

   \date{Received~~2009 month day; accepted~~2009~~month day}

\abstract{Dome A in the Antarctic plateau is likely one of the best astronomical observation sites on Earth. The first one of three Antarctic Survey Telescope (AST3-1), a 50/68 cm Schmidt-like equatorial-mount telescope, is the first trackable telescope of China operating in Antarctica and the biggest telescope located in Antarctic inland. AST3-1 obtained a large amount of data in 2012 and we processed the time-series parts. Here we present light curves of 29 variable stars identified from ten-day observations with AST3-1 in 2012, including 20 newly-discovered variable stars. 19 variables are eclipsing binaries and the others are pulsating stars. We present the properties of the 29 variable stars, including the classifications, periods and magnitude ranges in \emph{i} band. For 18 eclipsing binaries, the phased light curves are provided with the orbital period values well determined.
\keywords{techniques: photometric survey --- stars:
variables}
}

   \authorrunning{Li G., Fu, J. N. \& Liu, X. M. }            
   \titlerunning{VARIABLE STARS OBSERVED WITH THE AST3-1 TELESCOPE}  

   \maketitle

%
%
\section{Introduction}           
\label{sect:intro}

Observations of variable stars play a crucial role in the understanding of stellar evolutions, as well as stellar formations and properties. Study of variable stars can make an important contribution in astrophysical researches such as the age of Universe, the distance of galaxies, the composition of the interstellar medium and the behavior of the expanding Universe. However, continuous observation is impossible in mid-latitude observatories because of the alternation of day and night. Luckily, there is a continent where night can last for almost six month: Antarctica.

Antarctica is a continent with the highest average altitude, the driest climate and the lowest average temperature. After tests lasting almost two decades by astronomers from different countries, Antarctica continent is shown to be the best place for all kinds of astronomical observations (\cite{Storey2005}). Apart from its stable atmosphere, Antarctica continent possess excellent conditions as follows,

\begin{itemize}

\item Continuous observations: the latitude close to the South Pole is high enough for celestial objects to be observed almost continuously during the Antarctic nights (\cite{2008A&A...490..287S}).

\item Clear skies: at Dome C in Antarctica, there are 75\% to 90\% days without clouds (\cite{2008EAS....33...41R}).

\item Clean and dry atmosphere: the atmosphere in Antarctica is the cleanest on earth. In South Pole the amount of aerosol is one fifty of that in Hawaii (\cite{1995JGR...100.8967B}).
\end{itemize}

Furthermore, the pollutions like light pollution don't exist in Antarctica.
 
In order to take advantage of these remarkable observation conditions at Dome A, which is the highest place in the inland of Antarctica continent and considered as the best astronomical site in both optical/IR and sub-mm, the Chinese Small Telescope Array (CSTAR) was installed at Dome A in 2008 January. CSTAR undertook both site testing and science research tasks (\cite{2014ApJS..211...26W}). As a following plan, the trio Antarctica Survey Telescopes (AST3) were designed for wide-field survey in multi-band from Dome A, mainly to search for supernovae and extrasolar planets. 

\section{Facilities and Observations}

\label{sect:Obs}

\subsection{Facilities}

The first of three AST3 telescope (AST3-1) is a 50/68 cm Schmidt-like equatorial-mount telescope, with the focal ratio of F/3.73, equipped with complete pointing and tracking components. AST3-1 was mounted with a STA1600FT CCD camera, with 10560 $\times$ 10560 pixels of 9 $\mu$m, corresponding to 1 arcsec per pixel on the focal plane. Only half area was used to make exposure. The filter is SDSS \emph{i} (\cite{2014SPIE.9149E..2HW}). The CCD chip is cooled with TEC to take advantage of the extremely low air temperature at Dome A in the winter, which is about -60 $^{\circ}$C on average (\cite{2012SPIE.8446E..6RM}).

The Frame Transfer mode of CCD is adopted to terminate the exposure rather than the mechanical shutter. The CCD is divided into two parts, the frame store regions (the top and bottom quarters with 10560 $\times$ 2640 pixels) and the image area (the central half 10560 $\times$ 5280 pixels), corresponding to a 2.93 deg $\times$ 1.47 deg on the sky, which is the effective field of view of the AST3-1 (\cite{2012SPIE.8446E..6RM}).

\begin{figure}
\centering
\includegraphics[scale=0.3]{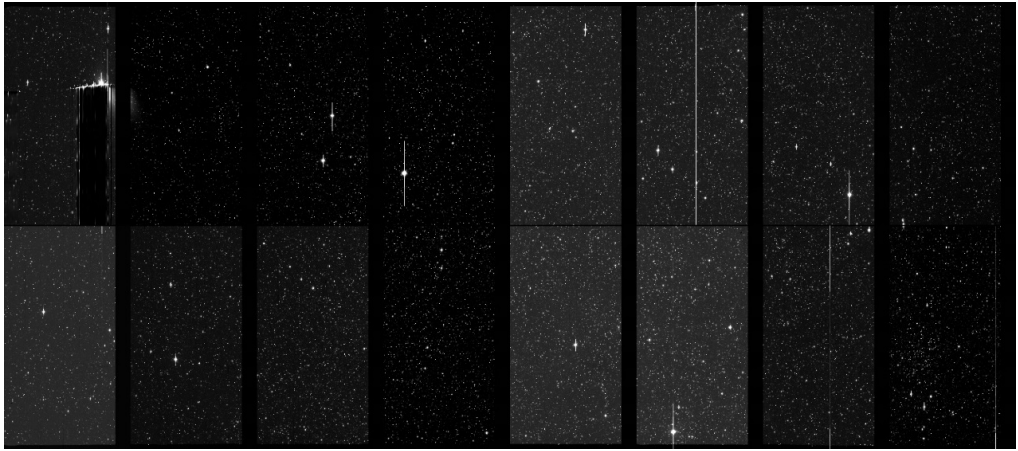}
\caption{A raw image of AST3-1. Note the 16 readout areas and its overscan areas. \label{fig1}}
\end{figure}

Figure~\ref{fig1} shows a raw image of AST3-1taken in 2012 with 5300$\times $12000 pixel$^2$, where one may see the image of stars and some bad pixels. There are sixteen readout amplifiers for the CCD chips to reduce the readout time. The area of one panel is 2640$\times$1320 pixel$^2$. Among the panels there are 8 vertical overscan areas whose size is 5300$\times$180 pixel$^2$ each and 1 horizontal overscan area whose size is 20$\times$12000 $pixel^2$.


\subsection{Observations}

AST3-1 was placed at Dome A by the 28$^{th}$ Chinese Antarctica Research Expedition (CHINARE) team in January of 2012 and carried out observations until May 8, 2012. In 2012, AST3-1 has in total observed 500 sky areas, corresponding to 2000 square degrees, and obtained more than 22000 images, including more than 3000 frames with exposure time of 60 seconds, and more than 4700 frames for LMC as well as 7000 images for the Galactic disk. The data collected with AST3-1 in 2012 were taken back to China by the 29$^{th}$ CHINARE team (\cite{2014SPIE.9154E..1TM}).

AST3-1 obtained a large amount of data in 2012 observation but the time-series data except LMC area is not much enough. This data was selected to process in order to find variable stars and it was list in table~\ref{tab1}.

\section{Data reduction}
\label{sect:data}

\subsection{Pre-reduction}

 The overscan was used as bias. After subtracting and cutting down the overscan, the image size was changed from 5300$\times$12000 to 5280$\times$10560. We ignored the effect of the dark current due to its low values. On the other hand, Ma (2014) gave a new method to deduce the dark currents of the AST3-1 data .

\begin{figure}
\centering
\includegraphics[scale=0.3]{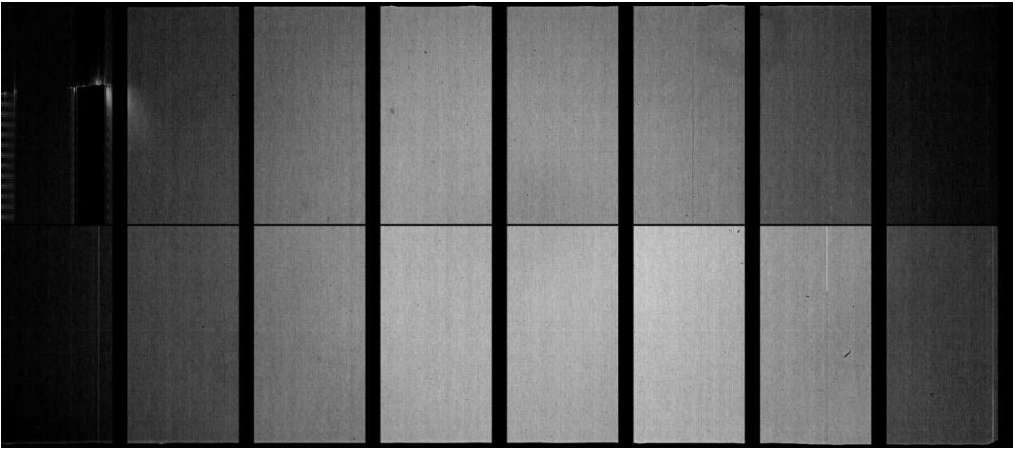}
\caption{The superflat \label{fig2}}
\end{figure}

\begin{figure}
\centering
\includegraphics[scale=0.3]{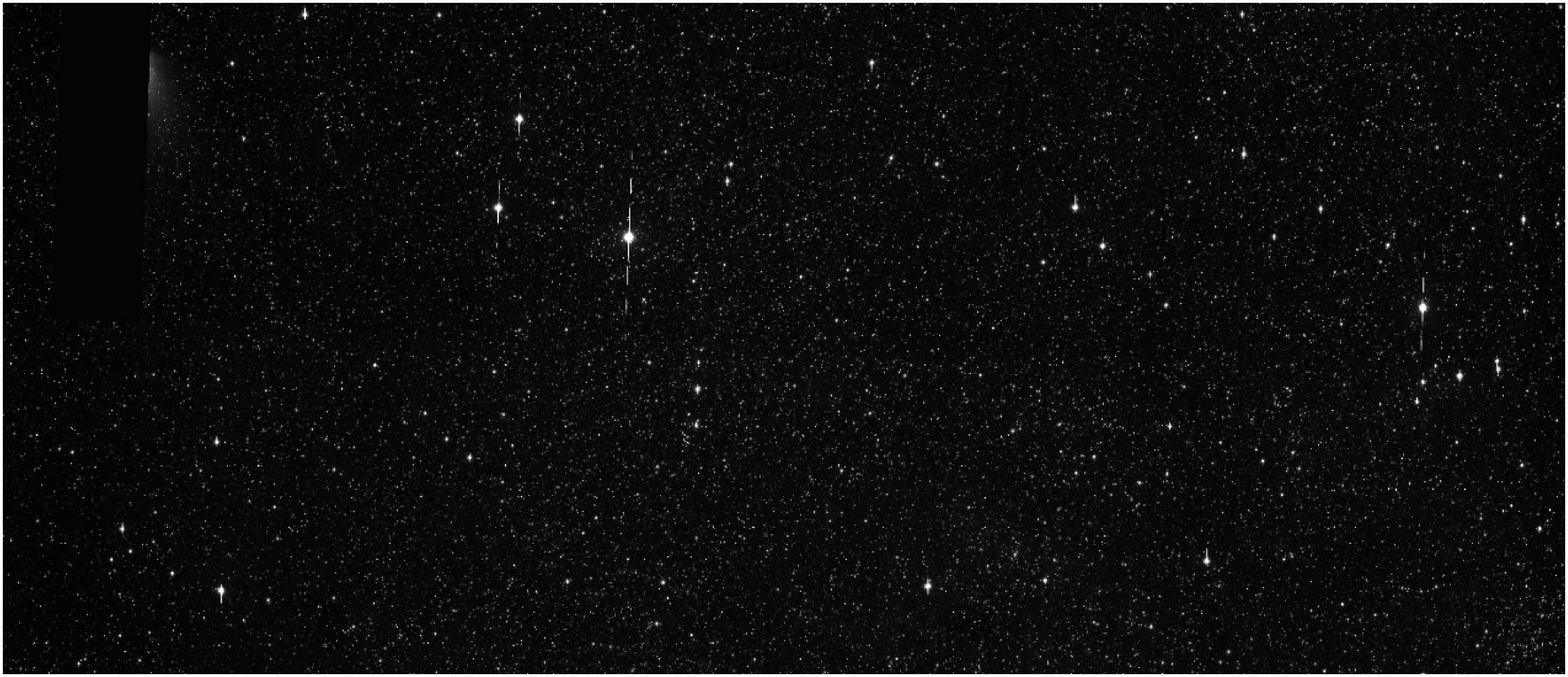}
\caption{An image after correction of bias and flat field \label{fig3}}
\end{figure}

As flat-field correction is a very necessary and critical step in data reduction, We selected 3763 images whose average ADU per pixel is larger than 5000 hence regarded as twilight images and combined them as a Super Flat, where the pixel values larger than twice of the background ones are replaced with the mean ADU value of the frame. Although the sky brightness gradient is minimum and stable, there is still a gradient of $\sim$ 1\% across the AST3\texttt{'}s  4.3 square degrees field-of-view (\cite{2014SPIE.9149E..2HW}). By combining numerous twilight images the gradient was averaged. After divided by the superflat, the images are ready for photometry. Figures~\ref{fig2} and~\ref{fig3} show the superflat and an image after the correction of bias and flat-field, respectively.

The data taken in 8$^{th}$ of April of 2012 show a good tracking capacity of AST3-1, as the images of stars on the frames kept stable during several hours\texttt{'} observations. However, the frames collected after 12$^{th}$ of April show problems of tracking. Picture shift and rotation as well as star trails become common. As a result we programmed an IDL script to align stars in different frames and removed bad data. Some gaps in the light curves which will be mentioned later are the consequence of the bad data.

\subsection{Photometry}

\label{sect:photometry}

For lists the datasets of AST3-1 collected in 2012, we used the command DAOFIND in IRAF to derive the coordinates of stars where the ADU values of the central pixel are larger than four times of the sky background ones to make sure a good photometric accuracy. The ADU values should be smaller than 30000, which is the limit of linearity of the CCD (\cite{2012SPIE.8446E..6RM}). For images whose exposure times are 60s each collected in 25th April, the number of stars satisfying this requirement is 12579. The result of flux calibration shows that the range of magnitude in \emph{i} band is approximately from $9^m$ to $15^m$. Aperture photometry is then used to derive light curves in order to detect the candidates of variables.

\section{Light Curves} 
 
After light curves obtained, some corrections were made before further analysis. From the data we noted that the clock of AST3-1 has a serious error which leads to the recorded time being the same for a group of images taken by continuous collection. In order to correct the time error, we assumed that for one group of images the time separations are the same hence the times of the images can be deduced. This assumption is proved to be reasonable by the phase diagrams of all the eclipsing binaries.

\begin{figure}
\centering
\includegraphics[scale=0.7]{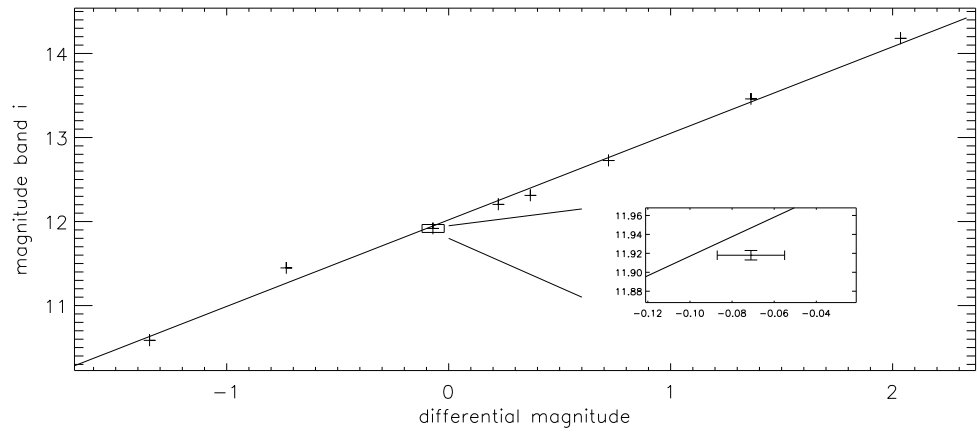}
\caption{Magnitude calibration. The linear fit shows a good magnitude calibration result. \label{fig4}}
\end{figure}

Figure~\ref{fig4} shows the magnitude calibration result. With the reference stars the linear fit leads to
$y=1.03(\pm 0.03)x+12.02(\pm 0.03)$, in which the related coefficient R² = 0.994, while $x$ means the differential magnitudes of eight check stars in \emph{i} and $y$ means magnitudes in \emph{i} from UCAC4 catalog (\cite{2013AJ....145...44Z}). 

\begin{figure}
\centering
\includegraphics[scale=0.6]{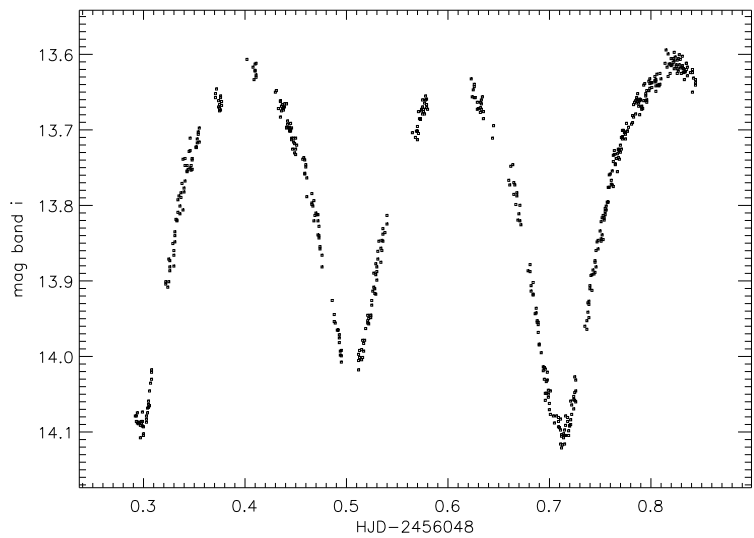}
\caption{Light curves of V23 in \emph{i} in 30$^{th}$ of April of 2012. \label{fig5}}
\end{figure}

Figure~\ref{fig5} shows the light curves of the star V23 observed in 30$^{th}$ of April of 2012, in which one can see that the light curves in even one night are discontinuous, due to bad data caused by either tracking components problems or bad pixels.
  
\begin{figure}
\centering
\includegraphics[scale=0.7]{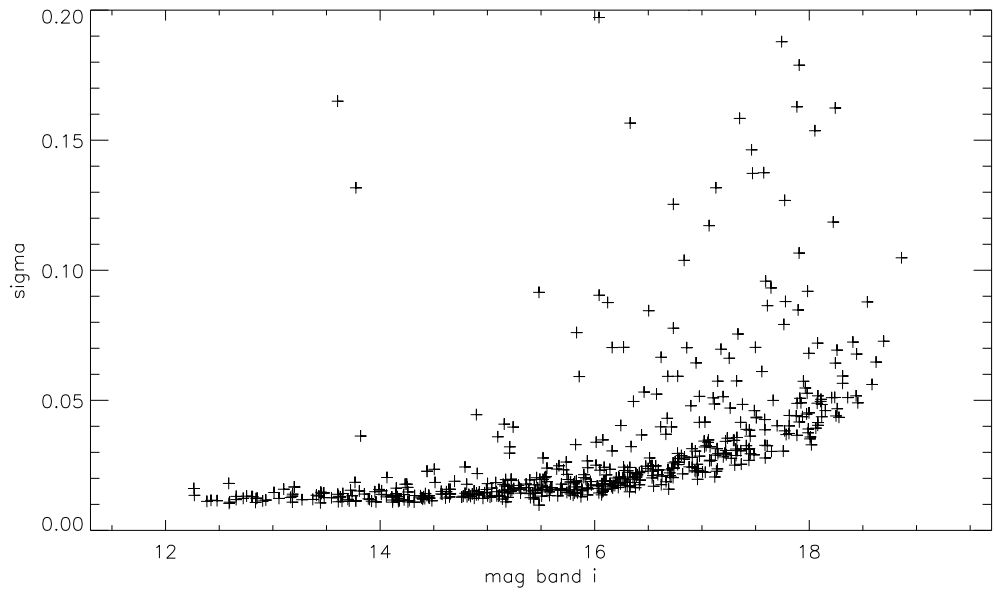}
\caption{Photometric accuracy versus magnitude in \emph{i} band when the aperture radius is 4 pixels.  \label{fig6}}
\end{figure}

  Figure~\ref{fig6} shows the diagram of the photometry accuracy versus the magnitude in \emph{i} band when the aperture radius was taken as 4 pixels. From Figure~\ref{fig6}, one may see that the photometry accuracy for most stars with magnitude in \emph{i} band between $12^m$ and $15^m$ is between $0^{m}.01$ and $0^{m}.02$. If the sigma for a certain star is larger than three times of the normal accuracy in its magnitude in Figure~\ref{fig6}, in the meanwhile no CCD problems were found by eyes, the star is regarded to be a variable.

\section{Results and Discussion}

\subsection{Variable stars}

Table~\ref{tab2} and Table~\ref{tab3} summarize properties of the 29 variable stars we detected. 20 ones are newly discovered in Table~\ref{tab2} after searching in the AAVSO website and the other ones have already been known which are listed in Table~\ref{tab3}. We present the information including name, coordinate in International Celestial Reference System (ICRS) from UCAC4 catalog (\cite{2013AJ....145...44Z}), classification, period or time scale and magnitude range in \emph{i}. Magnitude ranges are derived by using the polyfit to the extreme values. Three of the known variable stars were discovered by \emph{the Optical Gravitational Lensing Ex-periment (OGLE)} (\cite{2013AcA....63..115P}). Two were found by \emph{The All Sky Automated Survey(ASAS)} (\cite{2002AcA....52..397P}).

\begin{figure}
\centering
\includegraphics[scale=0.9]{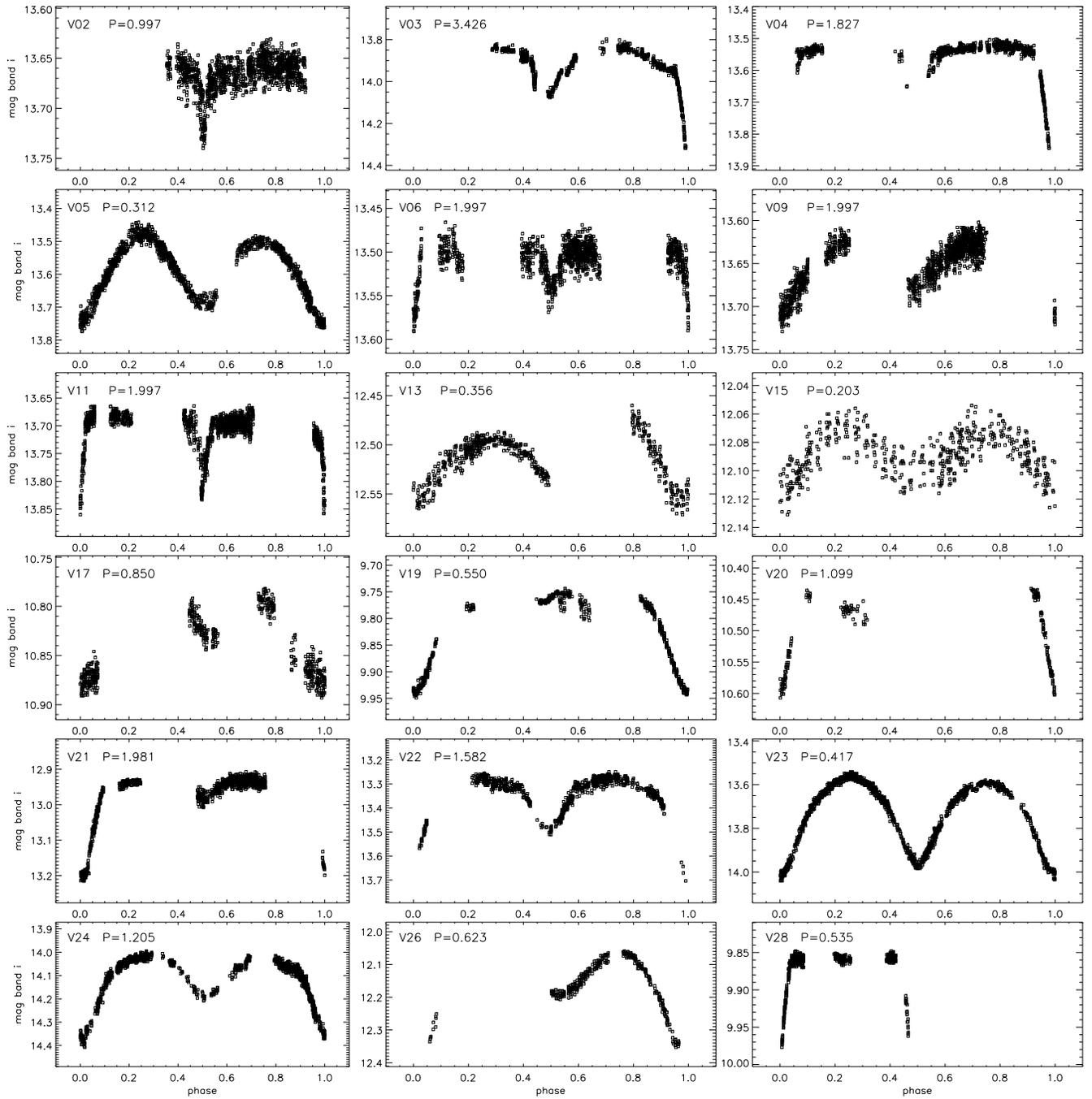}
\caption{Phased light curves of 17 eclipsing binaries among the 29 variable stars, listed in Tables~\ref{tab2} and ~\ref{tab3}  \label{fig7}}
\end{figure}

\begin{figure}
\centering
\includegraphics[scale=0.9]{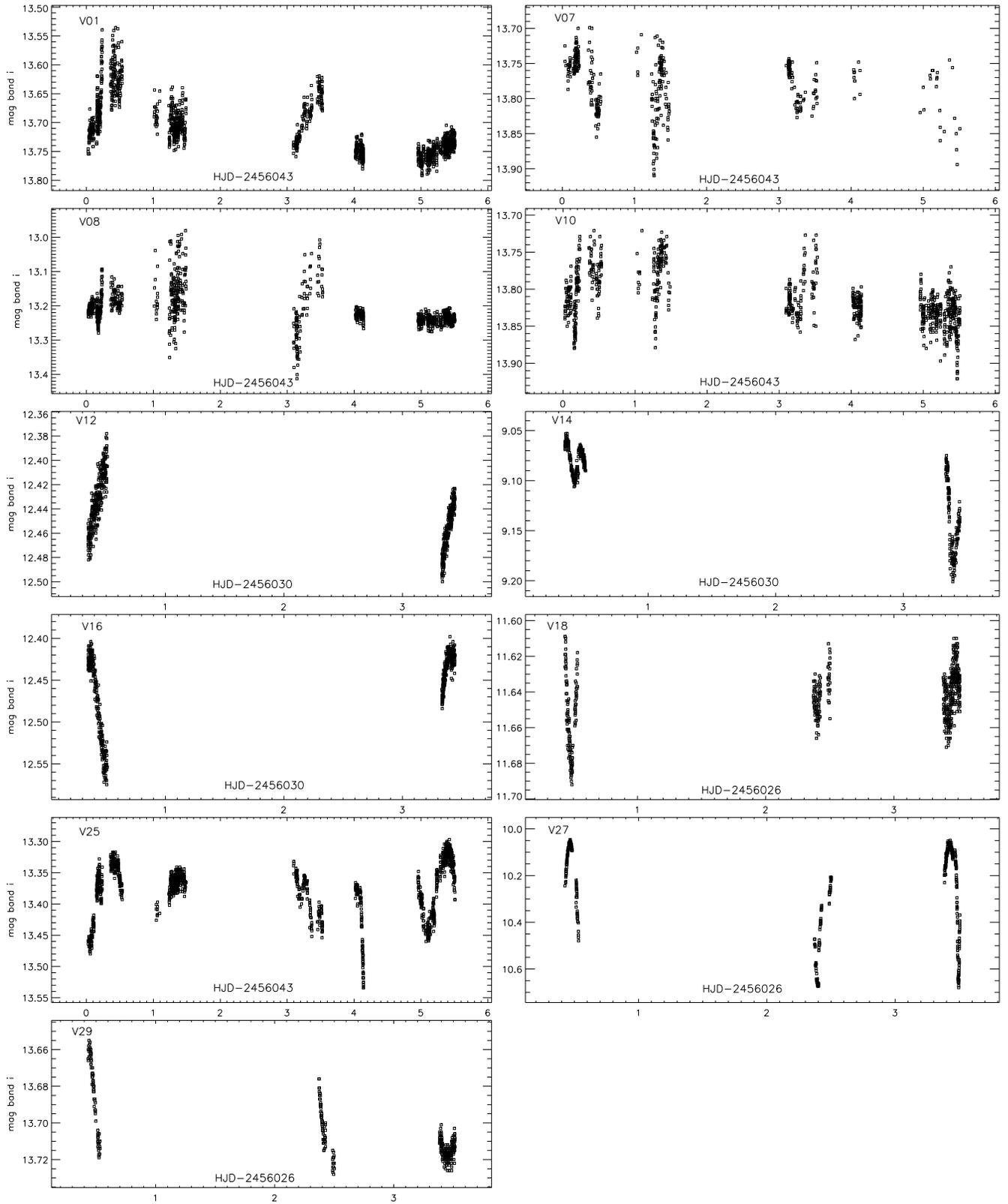}
\caption{Light curves of 12 vriable stars whose phased diagrams can not be made with certain values of periods. \label{fig8}}
\end{figure}

In order to deduce the orbital periods of eclipsing binaries from the light curves,  we analyzed the shapes of light curves to estimate the rough values of periods, then apply a ten-order polyfit to the light curves to optimize the periods hence obtain the phased light curves with the least-square fit. Figure~\ref{fig7} shows the phased light curves of the 18 eclipsing binaries among the 29 variable stars listed in Table~\ref{tab2} and Table~\ref{tab3}. The orbital period values are listed in the 5$^{th}$ column of Table~\ref{tab2} and ~\ref{tab3}. Figure~\ref{fig8} shows the light curves of the other 11 variable stars, which can not be combined to phase ones.

As light curves of 11 variable stars are not easy to be classified, we made a rough classification depending on their colour index and time scales, as well as the shapes of the light curves. We obtained their magnitude information from UCAC4 catalog, as listed in table~\ref{tab4}. Because of the variability of the luminosity, the effective temperatures determined with the colour index are not accurate.  We take three stars, as examples.
\begin{itemize}

\item[V01]: one can know from UCAC4 catalog that the magnitude of V01 is $m_J=13.269$ and $m_K=13.068$, so the colour index is $J-K=0.201$, which means that this star is F0 type with $T_{eff}=7030K$ (\cite{1998A&A...333..231B}). In the meanwhile, it is apparent that the increase is faster than decrease in light curve. Combining its time scale, it is possible for V01 to be an classical cepheid(\cite{1972vast.book.....S}).

\item[V08]: the magnitude is $m_B=13.580$ and $m_V=12.999$, so the colour index is $B-V=0.581$. The effective tempareture is hence $T_{eff}=6000K$ (\cite{1998A&A...333..231B}). This star is close to G0 type and likely to be a Cepheide \uppercase\expandafter{\romannumeral2} (\cite{1972vast.book.....S}) with the time scale of $\sim$ 1 day and the amplitude of $0^m.1$. Nevertheless, according to the shape of the light curves, we are not sure so it is classified as 'unknown' in Table~\ref{tab2}.

\item[V25]: the magnitude is $m_B=14.192$ and $m_V=13.748$, so the colour index is $B-V=0.444$. One can deduce the effective tempareture as $T_{eff}=6500K$ (\cite{1998A&A...333..231B}). A star between F0 and G0 with the time scale of 0.6 days and the amplitude of $0^{m}.13$ could be a $\delta$ Scuti star(\cite{1972vast.book.....S}). However, \cite{1963AnAp...26..153S} thought it as an eclipsing binary and gave the period of 1.676868 days. This inconsistency may be caused by the incomplete light curves in phase.

\end{itemize}

\subsection{Discussion}

\begin{figure}
\centering
\includegraphics[scale=0.7]{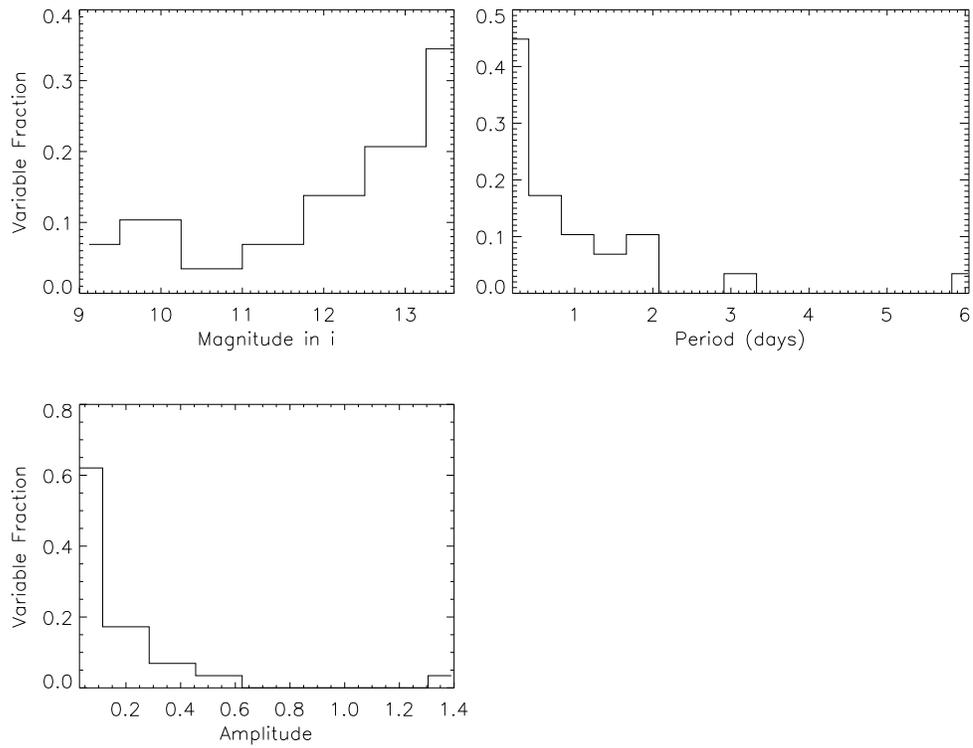}
\caption{The distributions of variable fraction versus the magnitude, period and amplitude. As to the magnitude distribution, one can see that the deeper the magnitude is, the larger the number of variable stars is. As for the period, one can see that the stars whith period shorter than 2 days occupy nearly 90\%. Note that the observation durations are shorter than one week for the data we processed. In terms of amplitudes, there is a dramatic drop in $0^m.1$ \label{fig9}}
\end{figure}

Figure~\ref{fig9} plots distributions of variable fraction versus the magnitude in \emph{i}, period and amplitude from the data of AST3-1 in 2012. We present a discussion and the photometric accuracy which are compared with the result from CSTAR (\cite{2015ApJS..218...20W}).

\begin{itemize}

\item Distribution of magnitude: From Figure~\ref{fig9}, one finds that the mean magnitude is $12^m.59$, and a peak appears at the magnitude larger than $13^m$, which is significantly deeper than CSTAR with mean magnitude of $\sim$ $10^m$.

\item Distribution of period: A jump is seen at 1 day. Since the observation data are shorter than one week, tt is then difficult to detect long-period variable stars and most detected periods are shorter than 2 days. By contrast, CSTAR can find the period up to 88.39days. AST3-1 also has such an ability with long-time observations.

\item Distribution of amplitude: Almost 60\% stars have small amplitude smaller than $0^m.1$, which is the same as that in CSTAR's result. The distribution proved that AST3-1 has the capacity to find stars with a slight varibility.

\item Photometric accuracy: As shown in Figure~\ref{fig6}, the photometric accuracies remain at the level of $\sim 0^m.01$ for stars of $12^m$ to $15^m$. As the contrast, the photometric accuracy of CSTAR rises from $0^m.05$ at $12^m$ to $0^m.2$ at $15^m$. This is believedto the larger aperture of AST3-1 than that of CSTAR.

\end{itemize}

In short, AST3-1 shows significant improvements in both deeper limit magnitudes and higher photometric accuracy.

\section{Conclusions}

Data observed with AST3-1 in 2012 are analyzed and 29 variable stars a are detected mong which 20 ones are newly discovered. For the stars with magnitudes in \emph{i} band in the range from $12^m$ to $15^m$, the typical accuracy of photometry is $0^m.01$. The magnitude calibration leads to a good result with the coefficient of 1.03. Properties of all the 29 variable stars are presented, including the classifications, periods and magnitude ranges in \emph{i} band. For 18 eclipsing binaries, the phased light curves are presented with the orbital period values well determined. Compared with the results from CSTAR, AST3-1 showed significant improvements in both limit magnitude and photometric accuracy. The data observed by AST3-1 in 2012 proved that observations by AST3-1 in Dome A can provide a unprecedented oppportunaties to the detection and study of variable stars.

\section{Acknowledgements}
The authors thank the support from the Scientific Research and Entrepreneurship Plan of Beijing College Students. JNF acknowledges the support from the Joint Fund of Astronomy of National Natural Science Foundation of China (NSFC) and Chinese Academy of Sciences through the Grant U1231202, and the support from the National Basic Research Program of China (973 Program 2014CB845700 and 2013CB834900).

\newpage

\newpage   
   
\newpage

\newpage

\newpage

\newpage

\newpage

\newpage

\newpage

\begin{table}
\begin{center}
\caption{Datasets we have already processed \label{tab1}}
\begin{tabular}{rrrrr}
\hline\hline
Date of observation & Name of field of view & $\alpha$ (ICRS) & $\delta$ (ICRS) &Number\\
\hline
$8^{th}$, $10^{th}$  and  $11^{th}$  Apr. &HD 143414 and WR star &240.9560&-62.6887&673\\
$12^{th}$  and  $15^{th}$ Apr. &HIP75377 & 231.0471 & -61.6771 & 618 \\
$25^{th}$  to  $30^{th}$ Apr. &transit & 163.9000 & -61.5000 & 1647\\
\hline
\end{tabular}
\end{center}
\end{table}

\newpage

\begin{table}
\centering
\begin{center}
\caption{Variable stars which are discovered by us. We list the name, coordinate, type, period or timescale (P/TS) in days and magnitude range in i (m range). If the phase is completed for an eclipsing binary, we give magnitude range in this format; primary/secondary max - primary/secondary min, otherwise we just give the range in: max - min. For the type, we indicate Eclipsing Binary(EB,EA,EC or EW), classical cepheid ($\delta$ Cep) and $\delta$ Scuti stars(DSct).\label{tab2}}
\begin{tabular}{rrrrrrr}
&\\
\hline\hline

Name & $\alpha$ (ICRS) & $\delta$ (ICRS) & Type & P/TS(d)  & m range\\

\hline

V01 & 10:52:32.93 & -61:45:54.9 & $\delta$ Cep & 3.0 & 13.63 - 13.76\\
V02 & 10:52:59.15 & -60:32:54.1 & EB & 0.997 &  13.68/no data - no data/13.88\\
V03 & 10:53:33.00 & -61:29:16.1 & EA & 3.426 &  13.32/no data - 14.84/no data\\
V04 & 10:53:55.55 & -61:28:13.8 & EA & 1.827 & 13.52/13.52 - 13.85/no data\\
V05 & 10:55:29.21 & -62:04:08.3 & EB & 0.312 & 13.48/13.50 - 13.74/13.70\\
V06 & 10:56:41.07 & -62:26:21.5 & EA & 1.997 & 13.50/13.50 - 13.54/13.60\\
V07 & 10:57:00.54 & -61:20:35.8 & unknown & 0.5 & 13.74 - 13.83\\
V08 & 10:57:10.97 & -61:58:34.1 & unknown & 1.0 & 13.15 - 13.25\\
V09 & 10:57:54.18 & -61:31:45.0 & EB & 1.997 & 13.63/13.64 - 13.70/13.68\\
V10 & 10:58:13.81 & -60:49:28.5 & unknown & 0.2 & 13.72 - 13.92\\
V11 & 10:58:52.16 & -61:57:31.9 & EA & 1.997 & 13.68/13.68 - 13.84/13.83\\
V12 & 15:21:57.46 & -62:09:15.7 & DSct & 0.3 & 12.41 - 12.50\\
V13 & 15:21:57.65 & -61:30:28.7 & EB & 0.356 & 12.48/12.49 - 12.55/12.54\\
V14 & 15:22:16.35 & -60:55:30.4 & DSct & 0.5 & 9.06/9.07 - 9.18/9.10\\
(HD 136194) \\
V15 & 15:22:25.31 & -61:21:06.7 & EB & 0.203 & 12.08/12.08 - 12.11/12.10\\
V16 & 15:22:55.53 & -61:16:38.2 & unknown & 0.4 & 12.42 - 12.55\\
V17 & 16:02:33.40 & -62:37:41.5 & EB & 0.850 & 10.80 - 0.88\\
(TYC 9040-2606-1) \\
V18 & 16:04:44.70 & -62:37:22.4 & DSct & 0.3 & 11.63 - 11.68\\
(TYC 9040-1073-1)\\
V19 & 16:05:03.11 & -62:05:33.7 & EA & 0.550 & 10.44/10.46 - 10.60/10.60\\
V20 & 16:05:18.75 & -62:07:24.6 & EA & 1.099 & 9.75 - 9.93\\
(HD 143660) \\
\hline
\end{tabular}
\end{center}
\end{table}

\begin{table}
\centering
\begin{center}
\caption{Variable stars which are discovered by us. We list the name, coordinate, type, period or timescale (P/TS) in days and magnitude range in i (m range). If the phase is completed for an eclipsing binary, we give magnitude range in this format; primary/secondary max - primary/secondary min, otherwise we just give the range in: max - min. For the type, we indicate Eclipsing Binary(EB,EA,EC or EW), classical cepheid ($\delta$ Cep) and $\delta$ Scuti stars(DSct).\label{tab3}}
\begin{tabular}{rrrrrrr}
&\\
\hline\hline

Name & $\alpha$ (ICRS) & $\delta$ (ICRS) & Type & P or TS  & m range\\

\hline
V21 & 10:52:05.33 & -61:20:31.7 & EA & 1.981 & 12.20/12.21 - 13.93/12.85\\
(OGLE-GD-ECL-04022)&   &   & EA & 1.977267 &   \\
V22 & 10:53:03.30 & -61:38:21.8 & EB & 1.582 & 13.27/13.27 - 13.71/13.50\\
(OGLE-GD-ECL-04249)&   &   & EC & 1.57794 &   \\
V23 & 10:53:45.48 & -61:15:35.0 & EC & 0.417 & 13.56/13.60 - 14.00/13.98\\
(OGLE-GD-ECL-04451) &   &   & EC & 0.41731592\\
V24 & 10:55:23.07 & -60:50:51.5 & EB & 1.205 & 14.02/no data - 14.36/14.17\\
(OGLE-GD-ECL-04845)&   &   & EC & 1.20580\\
V25 & 10:56:23.41 & -61:14:45.4 & DSct & 0.6 & 13.32 - 13.45\\
(V* IN Car) &  &   & EA & 1.676868\\
V26 & 15:59:05.37 & -62:40:36.2 & EB & 0.623 & 12.07/no data - 12.37/12.20\\
(ASAS J155905-6240.6) &   &   & EC & 0.62332\\
V27 & 15:59:06.40 & -63:17:49.4 & unknown & 1.1 & 10.07 - 10.67\\
(V* V336 TrA (\cite{2005IBVS.5600....1.})) &    &   & EW & 0.266768\\
V28 & 16:06:32.63 & -62:38:55.5 & EA & 1.634 & 9.86/9.86 - 9.99/no data\\
(ASAS J160633-6238.9	) &   &   & EA & 1.6343\\
V29 & 16:08:24.83 & -61:40:08.3 & EB & 1.0 & 13.66 - 13.72\\
(V* AD TrA (\cite{1931BHarO.884...10H})) &   &   & EB & 1.2604\\

\hline
\end{tabular}
\end{center}
\end{table}

\newpage

\begin{table}
\centering
\begin{center}
\caption{Information of variable stars from the literature whose types can not be deduced clearly. \label{tab4}}
\begin{tabular}{rrrrrrrr}
\hline\hline
Name & B(mag) & V(mag) & J(mag) & K(mag) & $B-V$ & $J-K$ & $T_{eff}$\\
\hline
V01 &   &   & 13.269 & 13.068 &   & 0.201 & 7030K \\
V07 &   &   & 12.096 & 11.296 &   & 0.800 & 3850K \\
V08 & 13.580 & 12.999 & 12.872 & 12.756 & 0.581 & 0.116 & 6000K \\
V10 &  & 13.021 & 12.194 & 11.491 &  & 0.703 & 4150K \\
V12 & 13.607 & 12.923 & 11.494 & 11.060 & 0.684 & 0.434 & 5750K \\
V14 & 10.215 & 9.743 & 8.649 & 8.189 & 0.472 & 0.460 & 6500K \\
V16 & 13.742 & 13.069 & 11.453 & 11.045 & 0.673 & 0.408 & 5660K \\
V18 & 12.397 & 11.897 & 10.864 & 10.612 & 0.500 & 0.252 & 6250K \\
V25 & 14.192 & 13.748 & 13.154 & 12.690 & 0.444 & 0.464 & 6500K \\
V27 & 11.511 & 10.696 & 8.897  & 8.301 & 0.815 & 0.596 & 5240K \\
V29 & 13.338 & 12.889 & 11.999 & 11.624 & 0.449 & 0.375 & 6400K \\
\hline
\end{tabular}
\end{center}
\end{table}

\newpage

\label{lastpage}

\end{document}